\documentclass[a4paper,journal]{IEEEtran}
\usepackage{amsmath,amsfonts}
\usepackage{algorithm}
\usepackage{array}
\usepackage[caption=false,font=normalsize,labelfont=sf,textfont=sf]{subfig}
\usepackage{textcomp}
\usepackage{stfloats}
\usepackage{url}
\usepackage{verbatim}
\usepackage{graphicx}
\usepackage[normalem]{ulem}
\usepackage{xcolor}
\usepackage{makecell}
\usepackage{enumerate}
\usepackage{cite}
\usepackage{xtab}
\usepackage{algorithm}
\usepackage{algpseudocode}
\usepackage{flushend}

\usepackage{array, multirow, booktabs, ragged2e}
\IEEEoverridecommandlockouts

\begin{document}

\title{Optimal Power Sharing for Hybrid Energy Storage Systems Based on Karush-Kuhn-Tucker Conditions \thanks{This work was developed in the framework of project HVDC4ISLANDS that has received funding by the CETPartnership, the Clean Energy Transition Partnership under the 2023 joint call for research proposals, co-funded by the European Commission (GA N.101069750) and with the funding organizations AEI (National State Research Agency, Spain), PtJ (MWIKE) (Projekträger Jülich/Forschungszentrum Jülich GmbH, Germany), GSRI (General Secretariat for Research and Innovation, Greece), SEAI (Sustainable Energy Authority of Ireland), FFG (Austrian Research Promotion Agency) and RCN (The Research Council of Norway).}}
\vspace{-0.2cm}
\author{Juan Diego Rios-Peñaloza, Milan Prodanovi\'c,~\IEEEmembership{Senior Member,~IEEE}, and \\ Javier Rold\'{a}n-P\'{e}rez,~\IEEEmembership{Senior Member,~IEEE}. \vspace{-0.32cm}} 

\markboth{ }%
{Shell \MakeLowercase{\textit{et al.}}: A Sample Article Using IEEEtran.cls for IEEE Journals}

\vspace{-0.2cm}

\maketitle

\begin{abstract}
Renewable energy power plants are increasingly expected to provide ancillary services to the grid.
Yet, the variability of sources such as solar photovoltaics requires additional operational flexibility to deliver these services.
In this context, hybrid energy storage systems (HESSs) offer an appropriate solution, because different technologies with complementary characteristics can be applied and leveraged to share the power demand and improve the overall performance.
Yet, a proper power sharing requires a detailed model of the storage elements.
This aspect has barely been studied in the literature, where most models consider constant efficiencies.
Moreover, most formulations are based on optimisation problems that are computationally too demanding to be solved in real time.
In this paper, a controller is proposed to minimise the losses of a HESS considering detailed power-dependent efficiency curves of each storage technology.
The power sharing method is based on the analytical verification of the Karush--Kuhn--Tucker (KKT) conditions, which makes it computationally efficient and suitable for real-world deployment.
The proposed controller is applied to a test case consisting of a 10~MW PV power plant with a HESS based on a lithium-ion battery and a redox-flow battery, each with 2.5~MW power and 5~MWh capacity. 
The main contributions are verified via numerical simulations performed in MATLAB/Simulink, whereas a real-time implementation deployed in OPAL-RT demonstrates the viability of the algorithm for real-time applications. 
\end{abstract}
\vspace{-0.1cm}

\section{Introduction}
\IEEEPARstart{T}{o} preserve the reliability of power systems, variable renewable energy sources are increasingly requested to provide grid services such as frequency reserve, voltage regulation, and black-start capability.
For this reason, renewable power plants are more frequently integrated with storage systems~\cite{Montanes2022}.
Depending on the application, different storage technologies are considered~\cite{Rana_RevPVESS_JES22,Teleke_RuleBased_TSE10,Bullich_APC_SolE17,NREL_PVBESS,FanIEEEOAJPE2023,Fan_TransEC_2023,Conte_MILP_TIA19,Andoni_Market_TIA16}.
Among them, lithium-ion batteries (LIBs) are the most common technology as a result of their excellent performance in terms of high power and energy densities and high efficiency.
In addition, they are characterised by a fast response that makes them suitable to provide fast reserves, and to operate in island mode during power outages. 
However, their performance and lifetime are strongly affected by large operating cycles and deep discharging.
On the other hand, redox-flow batteries (RFBs) constitute a cost-effective option that offers independent scalability of energy and power.
These batteries can withstand a very large number of cycles, have low self-discharge rates, and the electrolyte can be reused almost indefinitely.
Yet, their efficiency is in general lower than that of LIBs and is strongly dependent on the output power~\cite{Wang_JEnergyStorage_2023,CanizaresProceedings2023,TimmannIEEEAccess2023}. 
Single-technology storage systems are limited by their inherent drawbacks~\cite{Sutikno_RevPVESS_Access22}, whereas hybrid energy storage systems (HESSs) overcome these shortcomings by combining the complementary advantages of different storage technologies.
They constitute a relevant option to increase the flexibility of renewable power plants~\cite{Reveles2024} and improve the overall efficiency of the storage system~\cite{Batteries2023}.
Proper power management of these hybrid systems must consider the characteristics and limitations of each technology, aiming to maximise the performance while preserving the storage units from degradation. 

Many authors have developed power sharing strategies to dispatch the assets, following different objectives and approaches.
The simplest method divides the set-points equally between the resources~\cite{ChengTransSG2014}, limiting the flexibility of the controller.
More refined methods employ dynamic weighting factors that might depend, e.g., on the capacity of the resource~\cite{KimTransPE2010}.
In particular, the actual capacity of ESSs depends on their state of charge (SOC), and therefore the weighting factors are commonly chosen to account for the relation between the SOC and the total controllable reserve~\cite{ChengTransSG2014,GouveiaBook2014,MoranEEEIC2023,RiosPEDG}. 
Weighting factors can also be calculated using economic indices~\cite{IravaniTransSG2018,MadureiraREPQ2005} or more complex functions that take into account the state of the resource~\cite{Iravani_TransPS_PotentialFunc_2010}. 
Karush--Kuhn--Tucker (KKT) conditions are employed in~\cite{Liu_KKT_IET,Ge_KKT_Conf} to simplify bi-level optimisation frameworks into single-layer optimisation problems; however, they still necessitate numerical solvers for their solution.
KKT is employed also in~\cite{Bari_PrinceKKT} for the real-time dispatch of a microgrid, and its advantages are demonstrated in a real environment. 
All the above methods have been developed in the context of microgrids or virtual power plants, rather than power plants with storage.
Although there are many similarities, the primary objectives and hierarchical structure are usually different. 
The main objective of hybrid power plants is to follow a power reference (e.g., market schedule) and to dispatch power to the network, rather than to supply local loads.
Moreover, a hybrid power plant is usually less heterogeneous than a microgrid.
Therefore, the power plant controller (PPC) can use more detailed models to manage the assets and achieve optimal plant performance.

The main challenge of PPCs for hybrid storage-integrated systems is proper power sharing between storage units.
This is particularly relevant when the plant participates in the provision of ancillary services, such as frequency containment reserve or automatic frequency restoration reserve (aFRR).
In such cases, the time requirements might be tight and improper power sharing management might affect the operational life of the storage system~\cite{DebusschereFCR}.
Wang~\textit{et al.}~\cite{Guishi_PVHESS_TSTE14} employ a supercapacitor-RFB system to smooth the fluctuating output power of a photovoltaic (PV) power plant. 
The RFB is prevented from frequent cycling by setting fixed feasible operation points. 
Similarly, Ma~\textit{et al.}~\cite{Ma_AllocHESS_Access19} employ a supercapacitor-LIB hybrid storage to smooth PV power fluctuations.
The power sharing approach aims to reduce the total losses of the HESS, but employs constant loss factors.
Díaz-González~\textit{et al.}~\cite{Bullich_PVHESS_JES22} study a system composed of a lead-acid battery and a supercapacitor for peak power shaving and limitation of PV ramps. 
First, the set-points of the storage units are optimised to avoid degradation due to power peaks.
Then, a real-time controller limits PV power ramps, prioritising the use of the supercapacitors.
These works aim to minimise losses and, to some extent, storage degradation.
Yet, the models used are simplistic and the impact of these modelling assumptions can be significant~\cite{Debusschere}.
Recently, the same authors of this paper developed a real-time controller for a PV-LIB-RFB plant~\cite{RiosTSTE}.
The controller implements a short-range optimisation based on updated information to adjust the dispatch and comply with the schedules cleared in the day-ahead and reserve markets.
The use of detailed efficiencies enables improved management of the storage units.
However, the complexity of the optimisation problem may lead to excessive computational times that fail to comply with some tight ancillary service requirements.

\begin{figure*}[!t]
\centering
{\includegraphics[width=0.78\textwidth]
{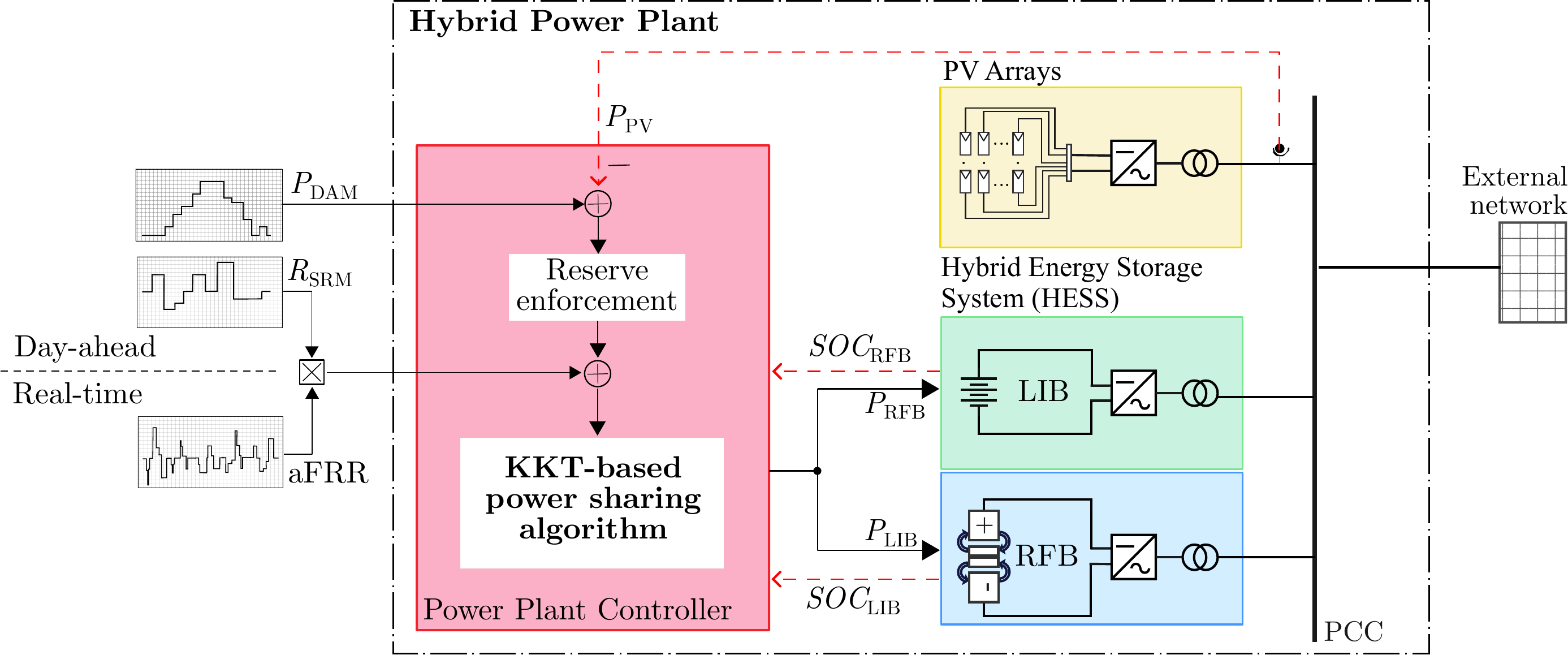}}
\caption{Diagram of the solar PV plant with HESS studied in this work.
(yellow) PV generators, (green) LIB, (blue) RFB and (red) proposed PPC.
}
\label{fig_rev.System}
\end{figure*} 

In this paper, a PPC is proposed for HESS-integrated solar PV power plants.
The controller implements a power sharing method that minimises the losses of the HESS by considering detailed power-dependent efficiency curves for each storage technology.
This method is based on analytical verification of the KKT conditions.
The problem is solved using a simple numerical root-finding method without the need for advanced optimisation solvers, making the algorithm computationally efficient.
Performance is tested on a 10~MW PV power plant equipped with a 5~MW HESS composed of a 2.5~MW LIB and a 2.5~MW RFB, each with a capacity of 5~MWh.
The system is simulated in MATLAB/Simulink.
A real-time implementation performed in OPAL-RT is used to demonstrate the feasibility of the algorithm in real scenarios. 

The remainder of this paper is structured as follows. 
Section~\ref{sec.SystemOverview} provides an overview of the plant controller and the application. 
Section~\ref{sec.PPC} describes in detail the PPC, the KKT conditions, and the storage modelling. 
Section~\ref{sec.Results} presents the simulation results and Section~\ref{sec.Conclusion} concludes the article.

\IEEEpubidadjcol

\section{System Overview}
\label{sec.SystemOverview}
\subsection{Application Description}
\label{sec.AppDescription}
Fig.~\ref{fig_rev.System} shows the hybrid power plant used in this work and the layout of the PPC.
The power plant consists of PV units and a HESS that includes a LIB and an RFB, all connected to the point of common coupling through power electronic converters.
The plant participates in the day-ahead market (DAM) and the secondary reserve market (SRM), providing aFRR capacity.
In the day-ahead stage, the power profile to be injected ($P_{DAM}$) and the reserve capacity offered for aFRR ($R_{SRM}$) are determined.
The main objective of the PPC is to guarantee that the plant follows the power profile cleared in the day-ahead stage, modified by the aFRR signal.
This signal is issued by the system operator during the operational day, such that the plant provides balancing energy according to the capacity reserved in that period.
\subsection{Overview of the Power Plant Controller}
Based on the previous description, the reference of the HESS is given by:
\begin{align}
P^*
&= 
P_{DAM}+aFRR \cdot R_{SRM}-P_{PV},
\label{eq.p_star}
\end{align}
where $P_{PV}$ is the power measured in the output of the array of PV units.

The ``Reserve enforcement'' block in Fig.~\ref{fig_rev.System} operates as follows.
The energy reserved in the SRM can be used only when balancing energy is required by the system operator, and cannot be used for trading energy in the DAM.
Assuming that the PV system operates at the maximum power point, the reserve capacity is provided exclusively by the storage units, which must maintain the energy corresponding to 15 minutes of such capacity~\cite{entsoe_afrr_2024}.
As a consequence, the effective SOC limits for the energy delivery in the DAM are more restrictive than the technical SOC limits of the storage units.
The ``Reserve enforcement'' block prevents the hybrid storage from violating these limits by setting the reference in the DAM to zero when they are reached.
For this purpose, the algorithm evaluates the reserve of capacity (ROC), which is equivalent to the SOC, but calculated in an aggregated way for the whole HESS.
This block allows the plant to prioritise the provision of balancing energy over complying with the DAM profile, rather than violating the committed reserve capacity.
The motivation for such a choice is the following. 
While violating any commitment (in the DAM or SRM) incurs an economic penalty, repeatedly failing to meet the reserve capacity could lead to the plant's exclusion from the ancillary service market.
This situation may occur, e.g., when PV production deviates significantly from the PV forecast used in the day-ahead scheduling.
Finally, the KKT-based power sharing algorithm determines the reference for each storage unit. 
The technical SOC limits are enforced within this stage.
This algorithm is described in detail in the next section.
\section{Hybrid Power Plant Controller}
\label{sec.PPC}
The objective of the power sharing strategy proposed in this paper is to minimise the losses of the HESS.
The problem is formulated analytically by deriving its KKT conditions and solved using a simple numerical root-finding method, without relying on advanced optimization solvers.
This approach makes the algorithm computationally efficient and suitable for real-time implementation.
\subsection{Optimisation Problem Definition}
The problem can be generalised as:
\begin{equation}
\begin{aligned}
\min \quad & \sum_{i=1}^N \ell_i, \\
\text{s.t.} \quad 
& \sum_{i=1}^N P_i = P^*, \\
& P_{i,\min} \le P_i \le P_{i,\max}, \quad i = 1,\dots,N,
\end{aligned}
\label{eq.minimisation}
\end{equation}
where $\ell_i$ represents the losses of the \textit{i}th storage unit over a period $\Delta t$, and depends on its own storage efficiency ($\eta_i$). 
$N$ is the total number of storage units. 
The constraints are defined such that the total power output meets the reference, ensuring that none of the units exceeds its power limits.
All powers are considered constant in the period $\Delta t$.

The following subsections describe the KKT conditions to accomplish~(\ref{eq.minimisation}), and the model of the losses used for the LIB and the RFB.
Finally, implementation aspects are explained.
\vspace{-0.2cm}
\subsection{KKT Conditions for Convex Resource Allocation}
\label{sec.KKTTheory}
Let us consider a generic power-dependent efficiency expression:
\vspace{-0.1cm}
\begin{align}
\eta_i 
&= 
f({P_i}).
\label{eq.eta_generic}
\end{align}
The efficiency curves are considered symmetric with respect to charging and discharging.
Therefore, $P_i$ denotes the power magnitude in the remainder of this section, i.e. $P_i \ge0$, and all loss and marginal loss functions are defined accordingly. 
The single storage loss function can be expressed as:
\begin{align}
\ell_i 
&= 
(1-\eta_i)P_i\Delta t,
\label{eq.loss_generic}
\end{align}
for a corresponding time step $\Delta t$.

An analytical KKT-based solution for~(\ref{eq.minimisation}) requires that each storage loss function is convex and that its marginal loss is continuous and strictly increasing over the operating range~\cite{WoodWollenberg}.
If these conditions are violated, the marginal-loss equalisation may not correspond to the global optimum, and numerical optimisation methods would be required.

At first glance, it is assumed that each loss function $\ell_i$ is convex in $P_i$ and each marginal loss $d\ell_i/dP_i$ is continuous and strictly increasing. 
In that case, the optimisation problem is convex and KKT conditions are sufficient for global optimality.
This assumption will be analysed in detail later.

The Lagrangian of the minimisation problem is simply:
\begin{align}
\mathcal{L}
&=
\sum_{i=1}^{N} \ell_i(P_i)
+
\lambda\!\left(
P^{*}-\sum_{i=1}^{N} P_i
\right)
+
\sum_{i=1}^{N} \underline{\mu}_i (P_{i}^{\min}-P_i)
\nonumber
\\
&+
\sum_{i=1}^{N} \overline{\mu}_i (P_i-P_{i}^{\max}),
\end{align}
where $\lambda$ is the Lagrange multiplier associated with the total power balance, and $\underline{\mu}_i$ and $\overline{\mu}_i$ correspond to the lower and upper power constraints, respectively.

In addition to the primal feasibility conditions (equalities and inequalities shown in~(\ref{eq.minimisation})) and the dual feasibility conditions ($\lambda$~$\ge$~$0$), KKT includes the stationarity condition~\cite{WoodWollenberg}:
\begin{equation}
\dfrac{d\ell_i}{dP_i}
-\lambda-
\underline{\mu}_i+
\overline{\mu}_i=0,
\label{eq.stationarity}
\end{equation}
and the complementary slackness:
\begin{align}
    \underline{\mu}_i (P_{i}^{\min}-P_i)&=0,
    \nonumber
    \\
    \overline{\mu}_i (P_i-P_{i}^{\max})&=0,
    \nonumber
    \\
    \underline{\mu}_i,\overline{\mu}_i &\ge 0.
\end{align}
The solution to the minimisation problem depends on $\lambda$:
\begin{equation}
P_i(\lambda) =
\begin{cases}
\left(\dfrac{d\ell_i}{dP_i}\right)^{\!-1}\!\!\!\!\!(\lambda),
\hspace{0.5em} \left.\dfrac{d\ell_i}{dP_i}\right|_{P_{i}^{\min}} \!\!\! < & \!\!\!\!\lambda < \left.\dfrac{d\ell_i}{dP_i}\right|_{P_{i}^{\max}}, \\[6pt]
P^{\min}_i, \hspace{1.5em} & \!\!\!\!\lambda \le \left.\dfrac{d\ell_i}{dP_i}\right|_{P^{\min}_i}, \\[6pt]
P_{i}^{\max}, \hspace{1.5em} & \!\!\!\!\lambda \ge \left.\dfrac{d\ell_i}{dP_i}\right|_{P_{i}^{\max}}.
\end{cases}
\label{eq.KKT_solution}
\end{equation}
The interior solution corresponds to the equalisation of marginal losses across all active units, i.e., $\lambda = d\ell_i/dP_i$ for all \textit{i}.
If a unit results outside its power limits, it is fixed to the corresponding bound, and the associated multiplier becomes active.
In this case, the marginal loss of that unit is not necessarily equal to the marginal value of the system ($\lambda$), but follows the inequalities indicated in~(\ref{eq.KKT_solution}).
\vspace{-0.1cm}
\subsection{Hybrid Storage Modelling}
\begin{figure}[!t]
\centering
\centering
\includegraphics[width=\columnwidth]{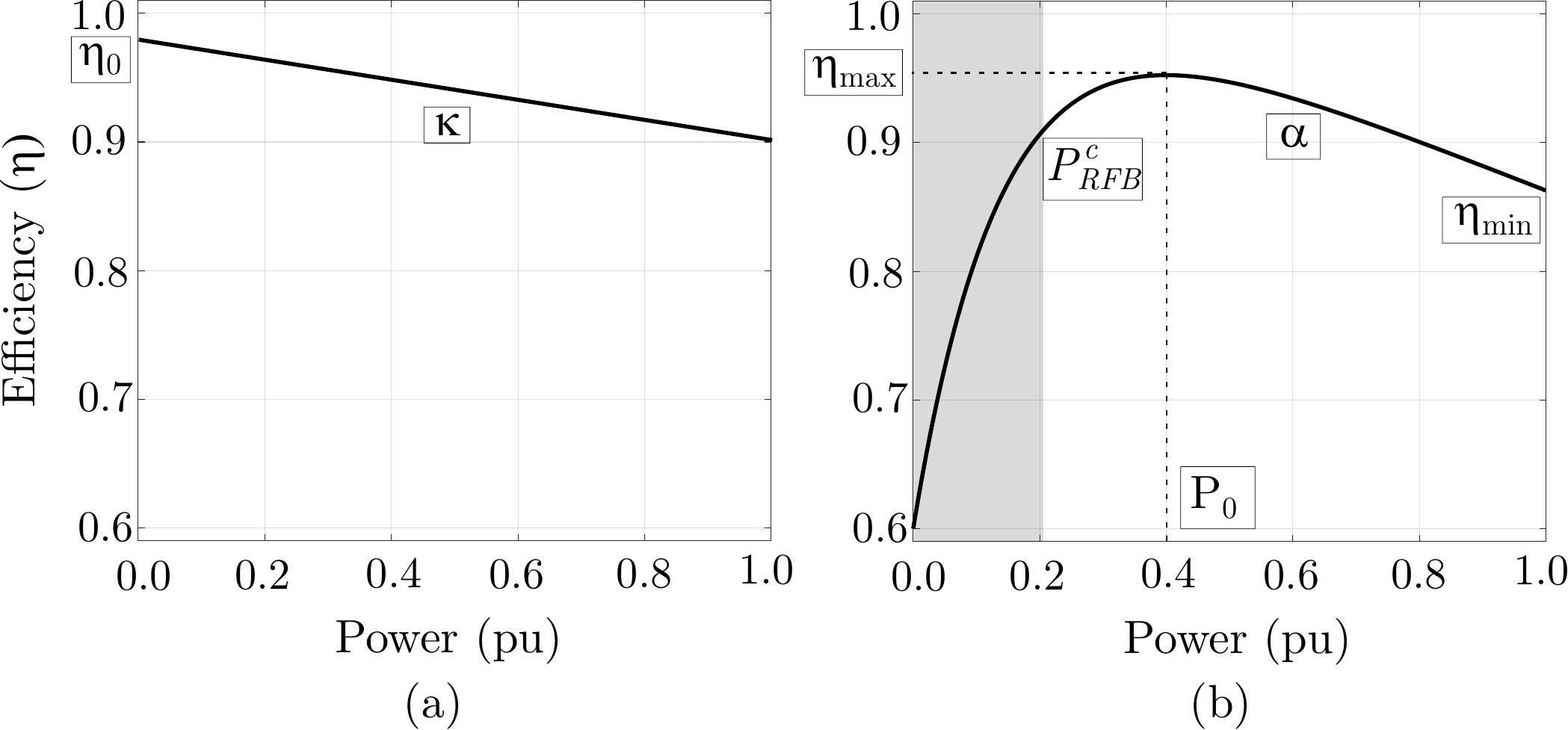}
\vspace{-.4cm}
\caption{Efficiency curves as functions of the power, for (a) LIB and (b) RFB.} 
\vspace{-.2cm}
\label{fig_rev.Efficiencies}
\end{figure}

Power injected/absorbed is the most influential parameter that affects battery efficiency, in addition to temperature, which is not considered in this paper for simplicity~\cite{FengChang_MathModel_ACS,Turker_ModelVRFB_ECM13,Su_ExperimentalLiB_JESt23,EnhancedLIB_Sakti_JPS2017}.
Fig.~\ref{fig_rev.Efficiencies} shows the curves representing the efficiency models employed in this paper, adapted from the experimental curves obtained in~\cite{FengChang_MathModel_ACS} for the RFB and~\cite{Su_ExperimentalLiB_JESt23} for the LIB.
The modelling details are explained in the following lines.
\subsubsection{Lithium-ion Battery}
For LIBs, the efficiency is modelled as a linear function that depends on the power absorbed/delivered by the battery:
\begin{align}
\eta_{LIB}(P)
&=
\eta_{\text{0}}^{} 
- 
\kappa \cdot {P}^{},
\label{eq.EfficLIB}
\end{align}
where $\bar{P}$ represents the absolute value of the power, $\eta_{\text{0}}^{}$ the maximum efficiency, and $\kappa$ a decay parameter.
Then, the power loss and marginal loss functions can be derived from~(\ref{eq.loss_generic}):
\begin{align}
\ell_{LIB}({P})
&=
(1-\eta_{\text{0}}+\kappa \cdot {P})\cdot {P},
\\
\dfrac{d\ell_{LIB}({P})}{d{P}}
&=
1-\eta_{\text{0}}+2\kappa \cdot {P},
\label{eq.MarginalLIB}
\end{align}
where the time step $\Delta t$ has been omitted for simplicity.
Fig.~\ref{fig_rev.LossFunctions} shows the power loss and marginal loss curves for the LIB, which are continuous and strictly increasing in $P$.
\subsubsection{Redox Flow Battery}
The efficiency of RFBs depends on the power delivered/absorbed, as in LIBs.
Nevertheless, for flow batteries, this dependence is strongly related to the losses associated with auxiliary systems such as the pumps and the cooling system.
Indeed, losses associated with pumping systems have a significant impact at low power levels~\cite{FengChang_MathModel_ACS}.
This is clearly represented in Fig.~\ref{fig_rev.Efficiencies}(b). 
The following equation is used to describe the RFB efficiency curve~\cite{RiosTSTE}:
\begin{align}
\eta_{RFB}(P)
&=
c_1-c_2 
\cdot 
e^{-\alpha \cdot P}-c_3 
\cdot 
P,
\label{eq.EfficRFB_P}
\end{align}
where $\alpha$ determines the shape of the efficiency curve.
Constants $c_1$, $c_2$ and $c_3$ depend on the maximum ($\eta_{\text{max}}$) and minimum ($\eta_{\text{min}}$) efficiencies, and the power at which efficiency is maximum ($P_o$):
\begin{align}
c_1
&=
\eta_{\text{min}}+c_2,
\label{eq.ConstantsRFB1}
\\
c_2
&=
(\eta_{\text{max}}-\eta_{\text{min}})/(1-(1+\alpha P_o) e^{-\alpha P_o}),
\\
c_3
&=
c_2\alpha e^{-\alpha P_o}.
\label{eq.ConstantsRFB3}
\end{align}

\begin{figure}[!t]
\centering
\centering
\includegraphics[width=0.935\columnwidth]{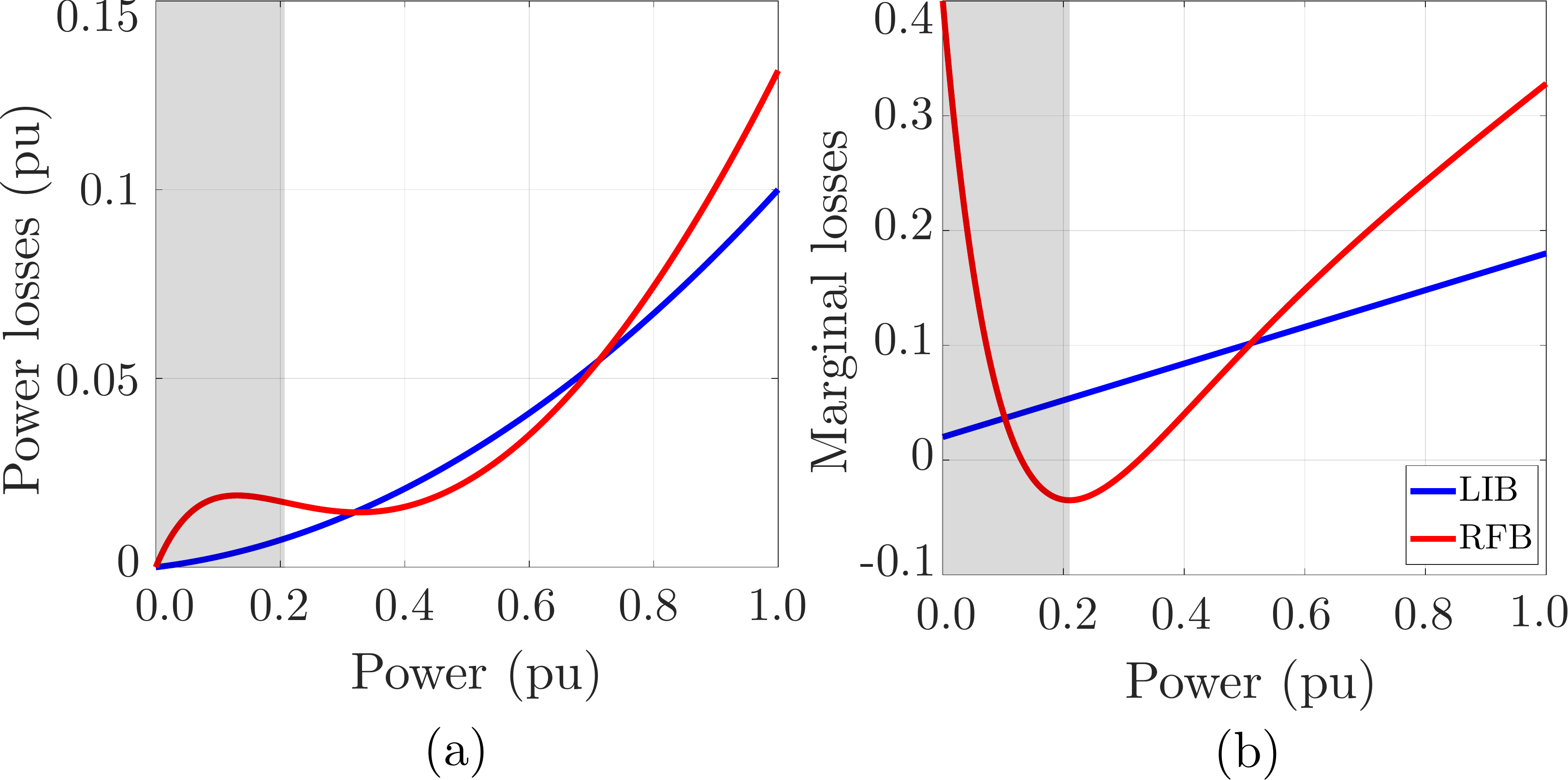}
\vspace{.0cm}
\caption{(a) Power loss functions and (b) marginal loss curves.} 
\vspace{-.3cm}
\label{fig_rev.LossFunctions}
\end{figure}

The RFB power loss and marginal loss are derived from~(\ref{eq.loss_generic}):
\begin{align}
\ell_{RFB}(P)
&=
(1-c_1+c_2 \cdot e^{-\alpha \cdot P}+c_3 \cdot P) \cdot P,
\\
\dfrac{d\ell_{RFB}(P)}{dP}
&=
1-c_1+c_2e^{-\alpha P}(1-\alpha P)+2c_3P.
\label{eq.MarginalRFB}
\end{align}
Fig.~\ref{fig_rev.LossFunctions} shows the power loss and marginal loss curves for the RFB.
The marginal loss is continuous in $P$, but not strictly increasing over the entire operational range.
Indeed, it decreases for very low values of $P$, namely for $P<P^{c}_{RFB}$, where $P^{c}_{RFB}$ features as the convexity power limit.
Its value can be obtained as the power corresponding to the minimum marginal loss, i.e., the minimiser of~(\ref{eq.MarginalRFB}).
The region where the marginal loss decreases is shown in grey in Fig.~\ref{fig_rev.Efficiencies}(b) and Fig.~\ref{fig_rev.LossFunctions}.
It can be intuited from Fig.~\ref{fig_rev.LossFunctions}(a) that the LIB alone is always preferred over this range of power, since its losses are significantly lower than those of the RFB.
In such a case, the non-monotonicity of the RFB marginal loss does not lead to ambiguity in the optimal dispatch.
In Section~\ref{sec.convexity}, a numerical validation will used to confirm this intuition.

\subsection{Optimisation Implementation}
A limitation of the formulation described in Section~\ref{sec.KKTTheory} arises from the commitment decisions.
Such formulation and the resulting expression~(\ref{eq.KKT_solution}) are useful to identify the optimal dispatch only when all units are active. 
When a storage unit is inactive, its losses are null; however, once it becomes active, it incurs a sudden positive loss.
This introduces a discontinuity in the marginal loss function at $P=0$, see Fig.~\ref{fig_rev.LossFunctions}(b).
This is analogous to the \textit{no-load cost} of classical generator scheduling~\cite{StrbacKirschen}.
This discontinuity at the activation boundary cannot be directly represented within the KKT framework illustrated in Section~\ref{sec.KKTTheory}.
Instead, the problem formulation must include commitment decisions as binary variables.
The minimisation problem is reformulated as:
\begin{equation}
\begin{aligned}
\min \quad & \textbf{u}\cdot\boldsymbol{\ell}, 
\end{aligned}
\label{eq.minimisation_2}
\end{equation}
where $\textbf{u}=[u_{LIB},u_{RFB}]$ is the state of the batteries (1 means active, 0 otherwise), and $\boldsymbol{\ell}=[\ell_{LIB},\ell_{RFB}]^T$ the vector of losses. 
Several methods, like dynamic programming and branch and bound, are commonly used to deal with these unit commitment problems.
However, since the number of units is small, a full enumeration approach can be easily implemented.

In addition to $\textbf{u}=[0,0]$, two different cases can be considered: i)~both units are active, i.e. $\textbf{u}=[1,1]$; ii)~one unit provides all the required power, i.e. $\textbf{u}=[1,0]$ or $\textbf{u}=[0,1]$. 
These cases are analysed in the following subsections. 

\subsubsection{Both Units Active}
As mentioned above, the formulation presented in Section~\ref{sec.KKTTheory} is useful when both units are active.
However, KKT conditions only give necessary conditions for a minimum, not a precise procedure to find that minimum, nor the inequalities that are binding.
The procedure to find the optimum usually involves trying various combinations of binding constraints until the KKT conditions are satisfied~\cite{WoodWollenberg,StrbacKirschen}.
With four binding variables ($\underline{\mu}_{i}, \overline{\mu}_{i}$, $i \in \{LIB, RFB\}$) and two possible values for each, the problem has $2^4$ $(16)$ possible combinations.
Yet, the problem can be simplified by substituting one of the variables in the equality constraint of the power balance, e.g., $P_{LIB}=P^*-P_{RFB}$.
What would have been expressed as two stationarity conditions (i.e.,~(\ref{eq.stationarity}) for each \textit{i}) becomes a single condition:
\begin{equation}
\begin{aligned}
\dfrac{d\ell_{LIB}(P^*-P_{RFB})}{dP_{RFB}}+
\dfrac{d\ell_{RFB}(P)}{dP_{RFB}}-\underline{\mu}_{RFB}+\overline{\mu}_{RFB}=0.
\end{aligned}
\label{eq.lagrangian_1D}
\end{equation}
If the binding constraints are off (i.e., the power is not at the boundaries),~(\ref{eq.lagrangian_1D}) becomes:
\begin{equation}
\begin{split}
c_2(1-\alpha P_{RFB}) e^{\alpha P_{RFB}} + (2c_3 + 2\kappa)\, P_{RFB} \\ + (\eta_0 - c_1 - 2\kappa P^*)= 0.
\label{eq.1D_KKT}
\end{split}
\end{equation}
The boundaries of the LIB must also be taken into account in this formulation, and they can be introduced as boundaries to the RFB as:
\begin{equation}
\small
P_{RFB} \in [\,\underbrace{\max(P^{\min}_{RFB},P^*-P^{\max}_{LIB})}_{\displaystyle P^{\min}_{RFB|LIB}},\underbrace{\min(P^{\max}_{RFB},P^*-P^{\min}_{LIB})}_{\displaystyle P^{\max}_{RFB|LIB}}\,].
\label{eq.limits_RFB}
\end{equation}
If the solution of~(\ref{eq.1D_KKT}) violates any of these boundaries, the power is forced to the relevant boundary.
Finally, the losses of this solution are computed as:
\begin{equation}
\ell
=
(1-\eta_{LIB})P_{LIB}+(1-\eta_{RFB})P_{RFB}.
\label{eq.tot_losses}
\end{equation}
It is important to note that, to guarantee the convexity of the problem, the condition $P^{\min}_{RFB} \ge P^{c}_{RFB}$ is required even if there are no technical constraints that limit the minimum power of the RFB.
\subsubsection{Single Unit Active}
If all the burden is taken by a single storage unit, the losses are computed as in~(\ref{eq.tot_losses}), where one of the two arguments will be null. 

\subsubsection{Optimisation Implementation}
The power sharing implementation considering the two cases is shown in Algorithm~\ref{alg:kkt_power_sharing}.
The case for which $P^*=0$ is trivial and is not included.
The following points are relevant for the algorithm implementation:
\begin{algorithm}[!t]
\caption{KKT-based power sharing}
\label{alg:kkt_power_sharing}
\small
\setlength{\baselineskip}{1.3\baselineskip}
\begin{algorithmic}[1]

\State \textbf{Offline inputs:} 
$\eta_{0}$, $k$, $\eta_{\min}$, $\eta_{\max}$, $\alpha$, $P_{o}$,
$P_{RFB(t)}^{\min}$, $P_{LIB}^{\min}$.

\State Compute the RFB efficiency parameters $c_{1}$, $c_{2}$, and $c_{3}$ using~(\ref{eq.ConstantsRFB1})--(\ref{eq.ConstantsRFB3}).
\State Compute $P^{c}_{RFB}$ as the minimiser of~(\ref{eq.MarginalRFB}).

\Statex \vspace{-0.4\baselineskip}\rule{\linewidth}{0.1pt}\vspace{-0.3\baselineskip}

\State \textbf{Online inputs:} 
$P^{*}$, $P_{RFB,LIB}^{\max}$.

\Statex Initialize the three candidate total losses:

\State $\ell_{\mathrm{tot}}^{(1,2,3)} \gets \infty$

\Statex \vspace{-0.6\baselineskip}\rule{\linewidth}{0.1pt}\vspace{-0.3\baselineskip}
\Statex \textbf{Case 1: Both units active:}
\Statex \vspace{-0.8\baselineskip}\rule{\linewidth}{0.1pt}

\State $P_{RFB}^{\min}=\max(P_{RFB(t)}^{\min},P^{c}_{RFB})$
\State Compute $P_{RFB}^{(1)}$ in $[P_{RFB}^{\min},P_{RFB}^{\max}]$ from~(\ref{eq.1D_KKT}) using \texttt{fzero}.
\State Enforce $P_{RFB}^{(1)} \in [P^{\min}_{RFB|LIB},\,P^{\max}_{RFB|LIB}]$ using~(\ref{eq.limits_RFB}).
\State $P_{LIB}^{(1)} \gets P^* - P_{RFB}^{(1)}$
\State $\ell_{\mathrm{tot}}^{(1)} \gets \ell_{RFB}(P_{RFB}^{(1)}) + \ell_{LIB}(P_{LIB}^{(1)})$

\Statex \vspace{-0.6\baselineskip}\rule{\linewidth}{0.1pt}\vspace{-0.3\baselineskip}
\Statex \textbf{Case 2: Single unit active:}
\Statex \vspace{-0.8\baselineskip}\rule{\linewidth}{0.1pt}

\If{$P^{*} \leq P_{RFB}^{\max}$}
    \State $P_{RFB}^{(2)} \gets P^{*}$, \quad $P_{LIB}^{(2)} \gets 0$,
    \State $\ell_{\mathrm{tot}}^{(2)} \gets \ell_{RFB}(P_{RFB}^{(2)})$
\EndIf

\If{$P^{*} \leq P_{LIB}^{\max}$}
    \State $P_{RFB}^{(3)} \gets 0$, \quad $P_{LIB}^{(3)} \gets P^{*}$
    \State $\ell_{\mathrm{tot}}^{(3)} \gets \ell_{LIB}(P_{LIB}^{(3)})$
\EndIf

\Statex \vspace{-0.4\baselineskip}\rule{\linewidth}{0.1pt}\vspace{-0.2\baselineskip}

\State $i^{*} \gets \arg\min_{i \in \{1,2,3\}} \ell_{\mathrm{tot}}^{(i)}$

\State \textbf{Return} $P_{LIB}^{(i^{*})},\,P_{RFB}^{(i^{*})}$

\end{algorithmic}
\end{algorithm}
\vspace{-0.3cm}
\begin{itemize}
\item The efficiency parameters and minimum technical power limits ($P^{\min}_i$) are constant system parameters introduced offline only once.
The convexity power limit for the RFB is also a constant that can be computed.
\item The inputs that are dynamic and are calculated online are the power reference ($P^*$ from~(\ref{eq.p_star})) and the maximum power ($P^{\max}_i$) for each storage unit.
The latter is considered a dynamic input because it depends on the SOC: if the SOC of any unit reaches its maximum (minimum), $P^{\max}_i$ is set to 0 for charging (discharging) operation; otherwise, it is set equal to the nominal power.
\item The three candidate total losses $\ell_{tot}^{(1,2,3)}$ are intialised, corresponding to $\left(\textbf{u}=\textbf{1}\right)$, $\left(\textbf{u}=[1,0]\right)$, and $\left(\textbf{u}=[0,1]\right)$, respectively.
The losses are computed for each case, and the lowest value is selected as the optimal dispatch.
\item To guarantee convexity when both units are active, the minimum power limit of the RFB is set as the larger of the technical minimum and the convexity limit.
\end{itemize}
\vspace{-0.1cm}
\section{Numerical Validation}
\label{sec.Results}
It is assummed the PV plant of Fig.~\ref{fig_rev.System} to have a capacity of 10~MWp, and the storage parameters are presented in Table~\ref{tab_rev.parameters}.
The system was simulated using MATLAB/Simulink.

\begin{table}[!t]
    \vspace{-0.45cm}
    \renewcommand{\arraystretch}{1.1}
    \caption{Parameters of the Storage Units}
    \vspace{-0.15cm}
    \centering
    \begin{tabular}{rll}
    \toprule[0.5pt]
    \textbf{Parameter} & 
    \textbf{LIB} & \textbf{RFB} \vspace{0.1cm}\\
    \hline
    Nominal power & 2.5~MW & 2.5~MW \\ 
    $P^{\max}\vert P^{\min}$ & 1\textbar 0~pu &1\textbar 0~pu \\ 
    Nominal capacity & 5~MWh & 5~MWh \\ 
    $SOC^{\max}\vert SOC^{\min}$ & 80\textbar 10\% & 95\textbar 5\% \\
    \multirow{2}{*}{Eff. parameters} & $\eta_0=0.98$ & $\eta_{\min}=0.6$, $\eta_{\max}=0.96$ \\
    & $\kappa=0.08$ & $P_0=0.4$, $\alpha=7$ \\
    \bottomrule[1.5pt]
    \end{tabular}
    \vspace{-0.2cm}
     \label{tab_rev.parameters}
 \end{table}

\vspace{-0.1cm}
\subsection{Convexity Limit Power}
\label{sec.convexity}
In this section, the intuition for which using the LIB alone is always preferred for $P_{RFB}<P^c_{RFB}$ is evaluated.

For the RFB used in this work, the convexity limit power $P^c_{RFB}$ is 0.21~pu.
First, the losses were calculated for the range \mbox{$P^*=[0,\ldots,1]$~pu} using only the LIB.
Then, the loss variation $\Delta\ell$ was calculated for different values of $P_{RFB}$ in the range \mbox{$[0, \ldots, P^*]$}.
From now on, the per-unit base value for the power reference was selected as the nominal power of a single storage unit, so the reference could range from -2 to 2~pu.

\begin{figure}[!b]
\centering
\vspace{-0.1cm}
\centering
\includegraphics[width=\columnwidth]{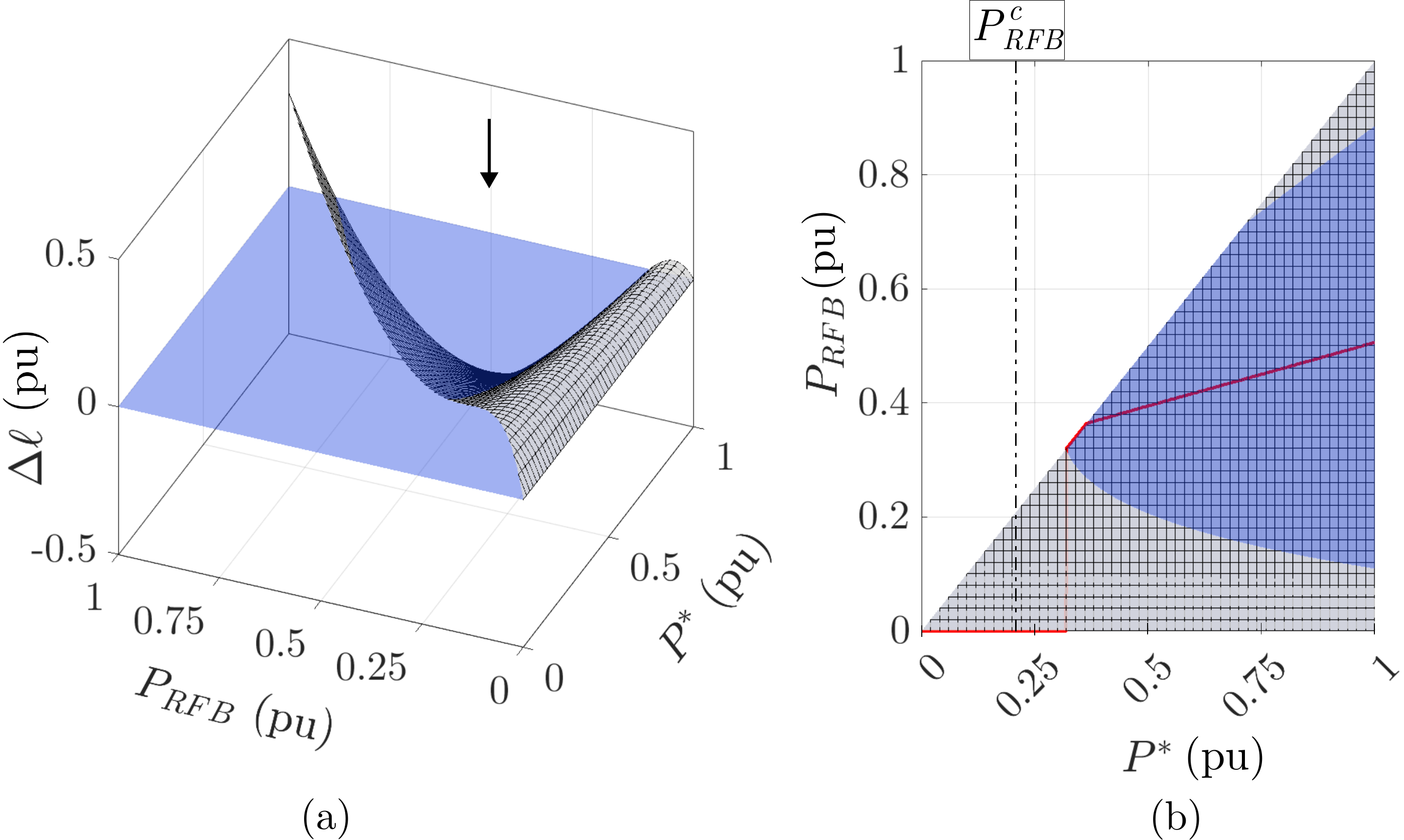}
\vspace{-.4cm}
\caption{(a) Losses variation ($\Delta \ell$) when substituting the LIB with the RFB: (a) 3D surface, and (b) cross-sectional view.} 
\vspace{-.2cm}
\label{fig_rev.rfb_3d}
\end{figure}

The results are shown in Fig.~\ref{fig_rev.rfb_3d}(a). 
The blue plane satisfies $\Delta\ell=0$ and helps differentiating the region in which losses decrease ($\Delta\ell<0$) or increase ($\Delta\ell>0$) compared to the case where only LIB is used.
Since the maximum possible value of $P_{RFB}$ is $P^*$, the surface appears only on one half-plane.
Fig.~\ref{fig_rev.rfb_3d}(b) shows the upper perspective of Fig.~\ref{fig_rev.rfb_3d}(a), allowing a better visualisation of the dependence of the losses on $P_{RFB}$, for different values of $P^*$ and $P_{RFB}$. 
The blue area shows the region in which losses decreased and the red line represents the minimum values of $\Delta\ell$, for different values of $P^*$.
It can be seen that it is inconvenient to activate the RFB until $P^*$ reaches a value greater than $P^c_{RFB}$ (0.23~pu, approximately).
By using these results, it can be concluded that, for the system under study, the non-monotonicity of the RFB marginal loss does not lead to ambiguity in the optimal dispatch.

The previous analysis suggests that a look-up table could be created offline, considering all possible values of $P^*$ and preallocating the power distribution.
While this is reasonable, the algorithm maintains a level of generalisation and flexibility to different parameters. 

\vspace{-0.1cm}
\subsection{Numerical Simulations}
In this section, the results of numerical simulations performed in MATLAB/Simulink to test the power sharing algorithm are provided.
To reduce the computational time, the simulation was scaled so that five seconds correspond to one minute of real operation.

\subsubsection{Dynamic Allocation}
\begin{figure}[!t]
\centering
{\includegraphics[width=\columnwidth]
{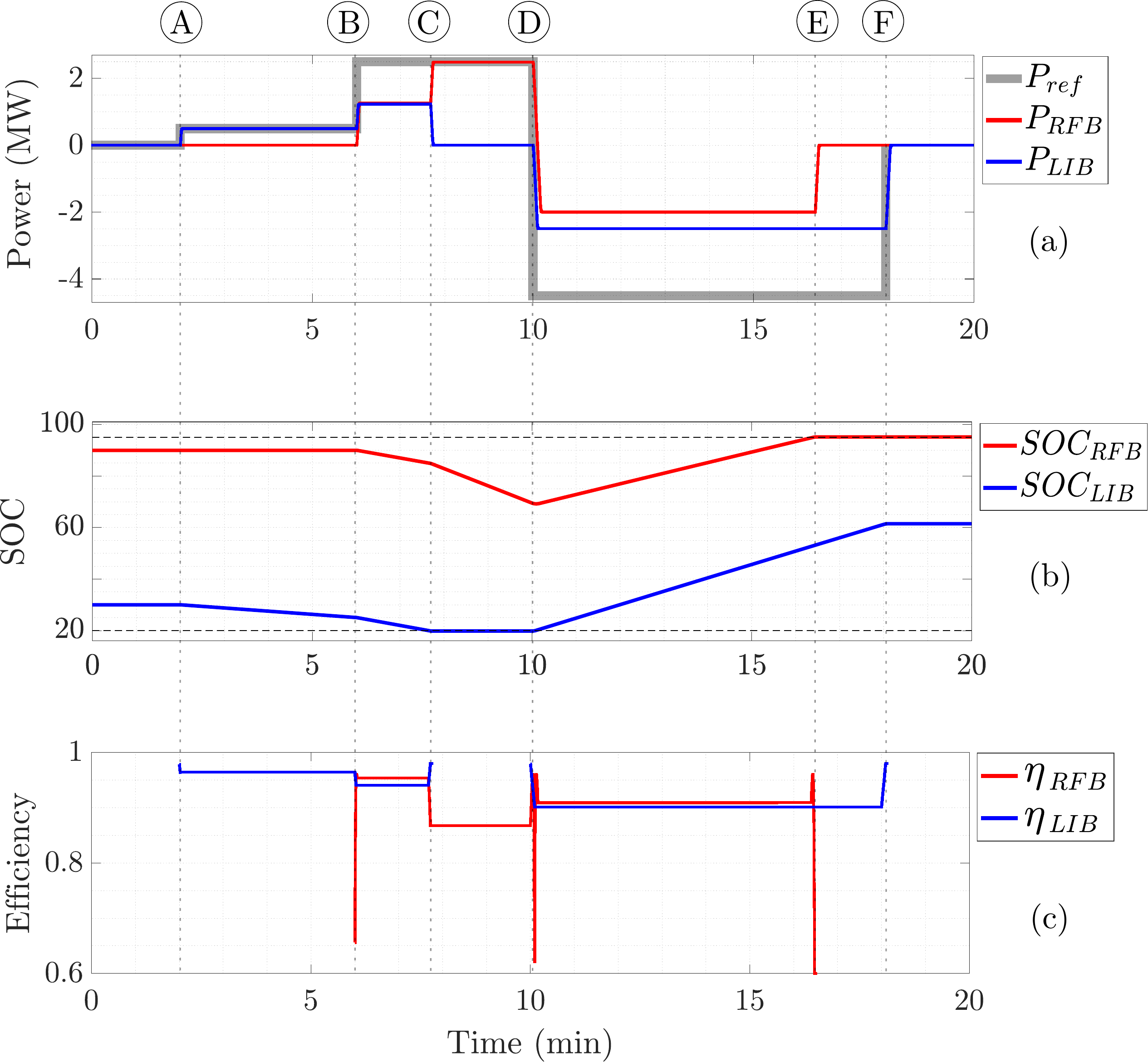}}
\vspace{-0.3cm}
\caption{(a) Power references for the HESS and for the single units; (b) SOC and (c) efficiencies of the storage units.}
\label{fig_rev.fig1y2}
\vspace{-0.3cm}
\end{figure}

First, a simulation was performed using a synthetic power reference profile to force the storage system to operate in different regions.
PV generation was considered null so that the plant output comes solely from the storage units.
The initial SOC of RFB and LIB was 92.5\% and 30\%, respectively, and each unit was forced to reach the maximum and minimum SOC limits during the simulation.
The results are shown in Fig.~\ref{fig_rev.fig1y2}. 
The power reference for the HESS is shown in Fig.~\ref{fig_rev.fig1y2}(a), as well as the power sharing between individual units; their SOC and efficiency values are shown in Fig.~\ref{fig_rev.fig1y2}(b) and Fig.~\ref{fig_rev.fig1y2}(c), respectively.

In \raisebox{.5pt}{\textcircled{\raisebox{-.9pt}{\small A}}} the reference was set to 0.2~pu (lower than $P^c_{RFB}$), and all the contribution is provided by the LIB. 
In \raisebox{.5pt}{\textcircled{\raisebox{-.9pt}{\small B}}} the reference was set to 0.8~pu, so both units were active. 
As can be seen in Fig.~\ref{fig_rev.fig1y2}(b), in \raisebox{.5pt}{\textcircled{\raisebox{-.9pt}{\small C}}} the SOC of the LIB reached its minimum, so the entire burden was taken by the RFB.
In \raisebox{.5pt}{\textcircled{\raisebox{-.9pt}{\small D}}} the total reference is set to -1.6~pu so both units charge.
In \raisebox{.5pt}{\textcircled{\raisebox{-.9pt}{\small E}}} the SOC of the RFB reached its maximum, but the LIB was already operating at its rated capacity.
Finally, in \raisebox{.5pt}{\textcircled{\raisebox{-.9pt}{\small F}}} the reference was set to zero. 

Fig.~\ref{fig_rev.fig1y2}(c) shows the efficiencies of the storage units when both were active. 
In particular, the efficiency for the RFB was always relatively high, except when it was forced to output the entire reference due to the SOC limits of the LIB.

\subsubsection{Daily operation}
The proposed power sharing algorithm was tested in a full-day operation, with the algorithm executed every 200~ms of real operation time.

\begin{figure}[!t]
\centering
\centering
\includegraphics[width=0.9\columnwidth]{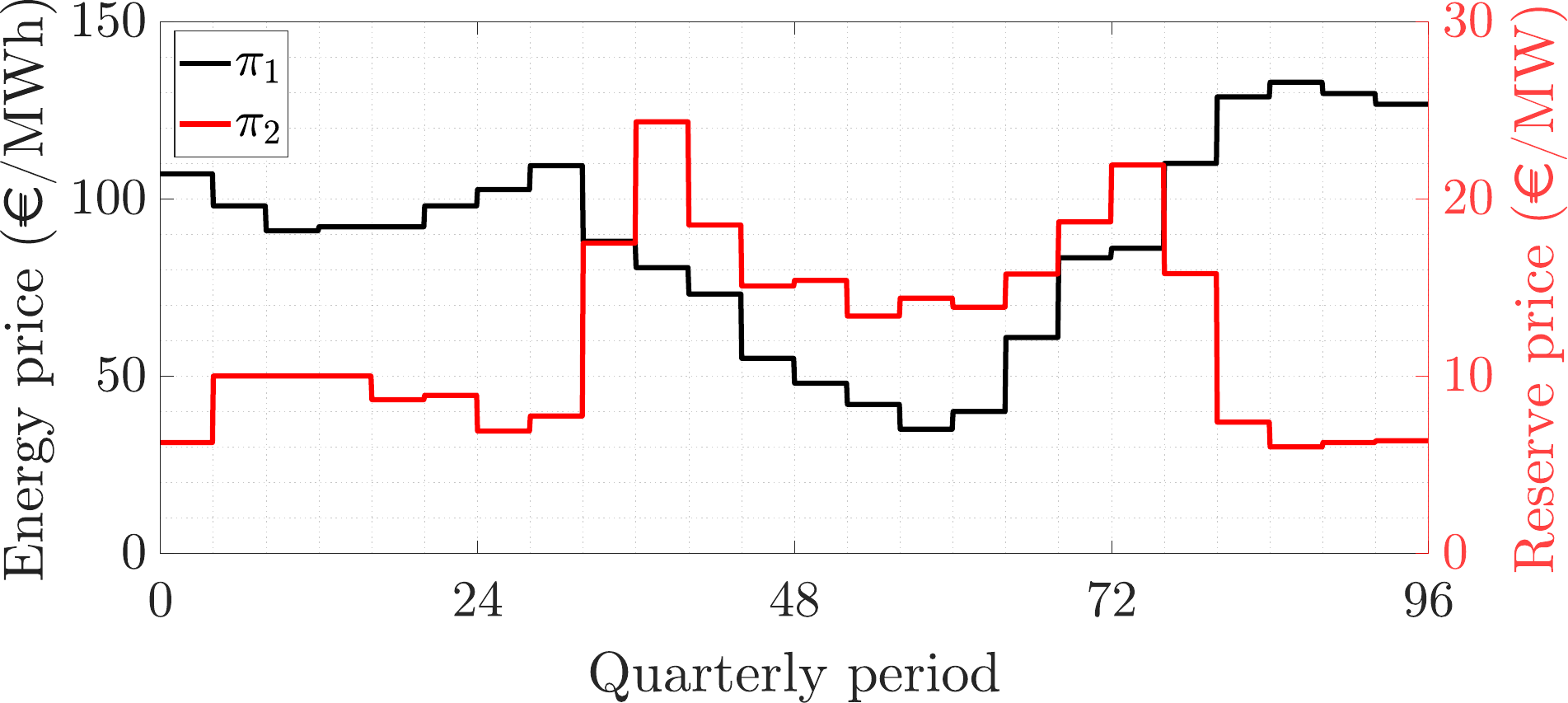}
\vspace{-.3cm}
\caption{Energy ($\pi_1$) and reserve ($\pi_2$) prices.} 
\vspace{-.2cm}
\label{fig_rev.prices}
\end{figure}

\begin{figure}[!t]
\centering
\centering
\includegraphics[width=\columnwidth]{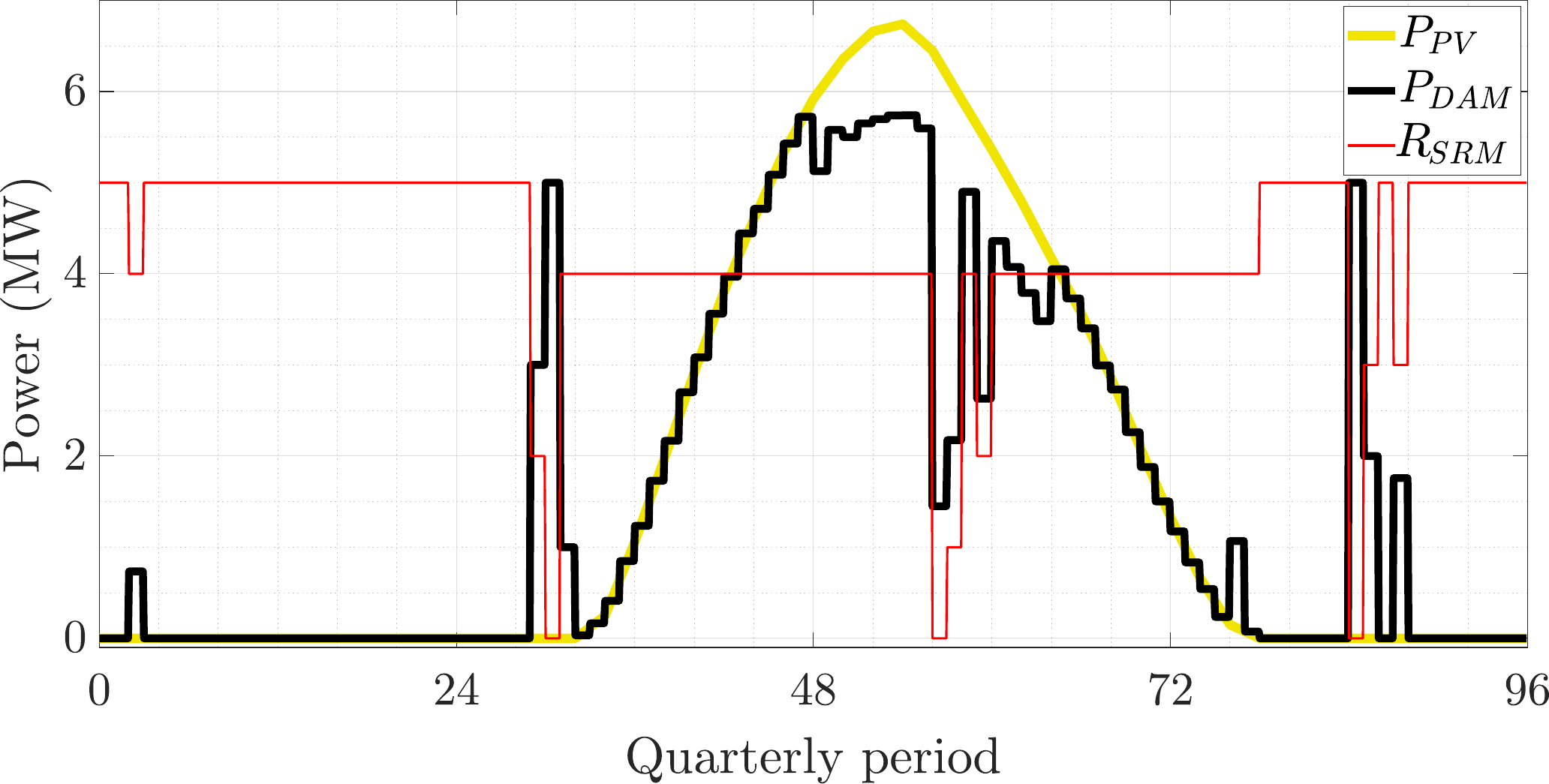}
\vspace{-.3cm}
\caption{Power and reserve profiles committed to the day-ahead and secondary reserve markets.} 
\vspace{-.2cm}
\label{fig_rev.DAM}
\end{figure}

The plant was scheduled to consider participation in DAM and SRM, following the strategy reported in~\cite{RiosTSTE}.
Fig.~\ref{fig_rev.prices} shows the energy and reserve prices considered for the DAM and SRM scheduling.
These prices were based on historical data obtained from \textit{Red Eléctrica de España}~\cite{REE_Esios}, the Spanish transmission system operator.
Although the prices were constant in each hour, scheduling was performed quarter-hourly, in line with the new market resolution. 
Fig.~\ref{fig_rev.DAM} shows the main results of the day-ahead scheduling, namely: the PV forecast generation profile ($P_{PV}$); the power committed by the power plant in the energy DAM ($P_{DAM}$); and the reserve profile ($R_{SRM}$).
All these signals have a one-minute time granularity.
To avoid the influence of prediction errors and assess the performance of the proposed controller individually, in this section it is assumed that PV generation followed the predicted profile.
For the same purpose, the ``Reserve enforcement'' block was deactivated.
After this section, the effect of these elements will be considered and studied.

\begin{figure}[!t]
\centering
\centering
\includegraphics[width=\columnwidth]{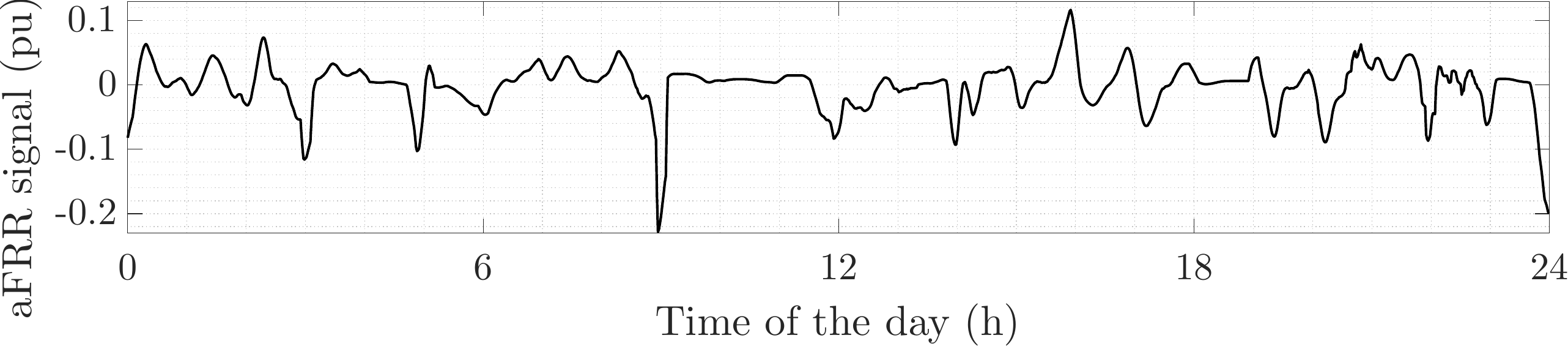}
\vspace{-.5cm}
\caption{Reserve activation signal.} 
\vspace{-.3cm}
\label{fig_rev.afrr_signal}
\end{figure}

\begin{figure}[!t]
\centering
\centering
\includegraphics[width=\columnwidth]{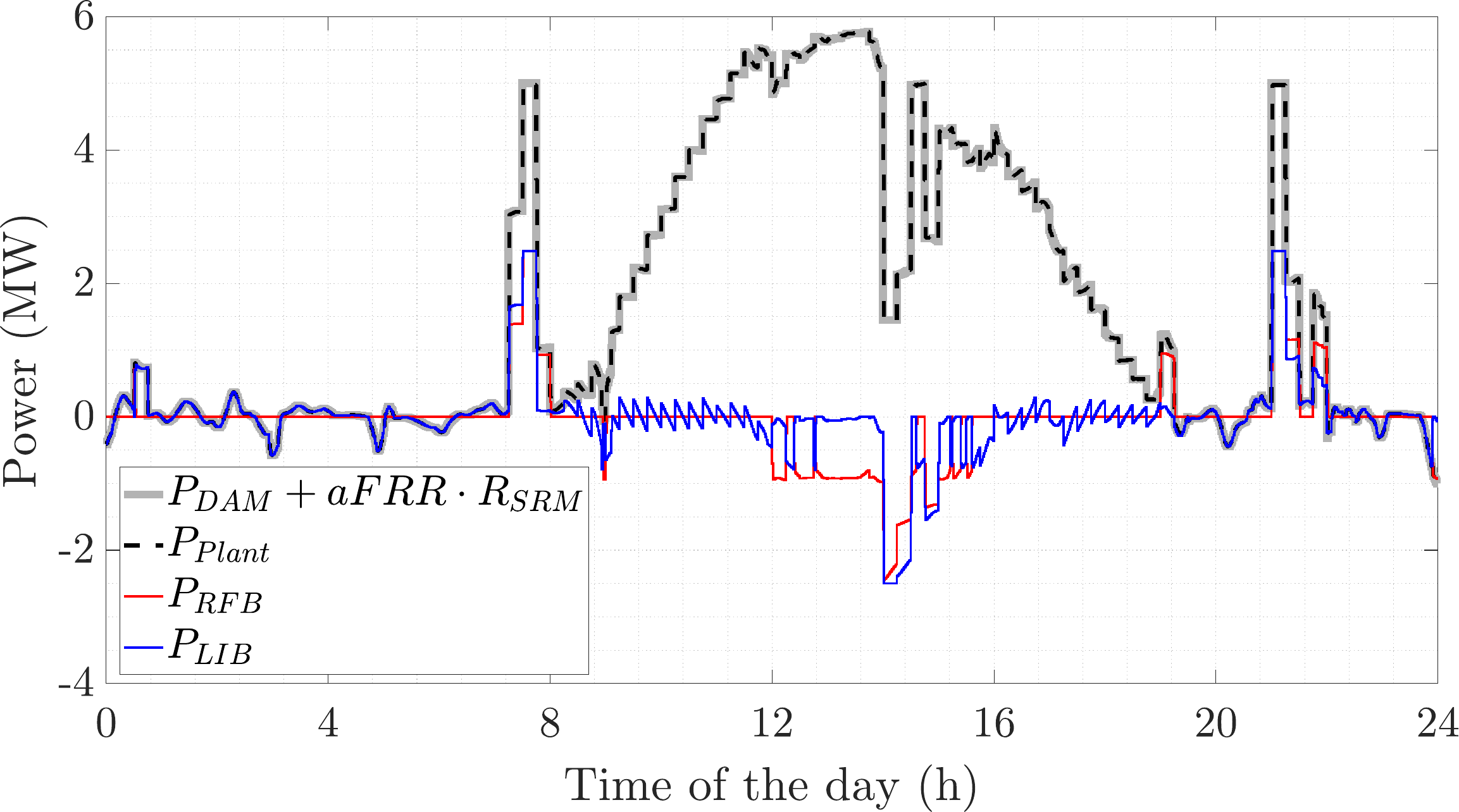}
\vspace{-.5cm}
\caption{Power set-points of individual storage units and total plant power.} 
\vspace{-.3cm}
\label{fig_rev.ProfilesResults}
\end{figure}

Fig.~\ref{fig_rev.afrr_signal} shows the activation signal for the reserve, in per unit of the reserve capacity committed.
The results of the operation of the PV plant are shown in Fig.~\ref{fig_rev.ProfilesResults}.
They show the reference profiles of the storage units and the correspondence between the power output of the plant and the committed power.

\begin{figure}[!t]
\centering
{\includegraphics[width=\columnwidth]
{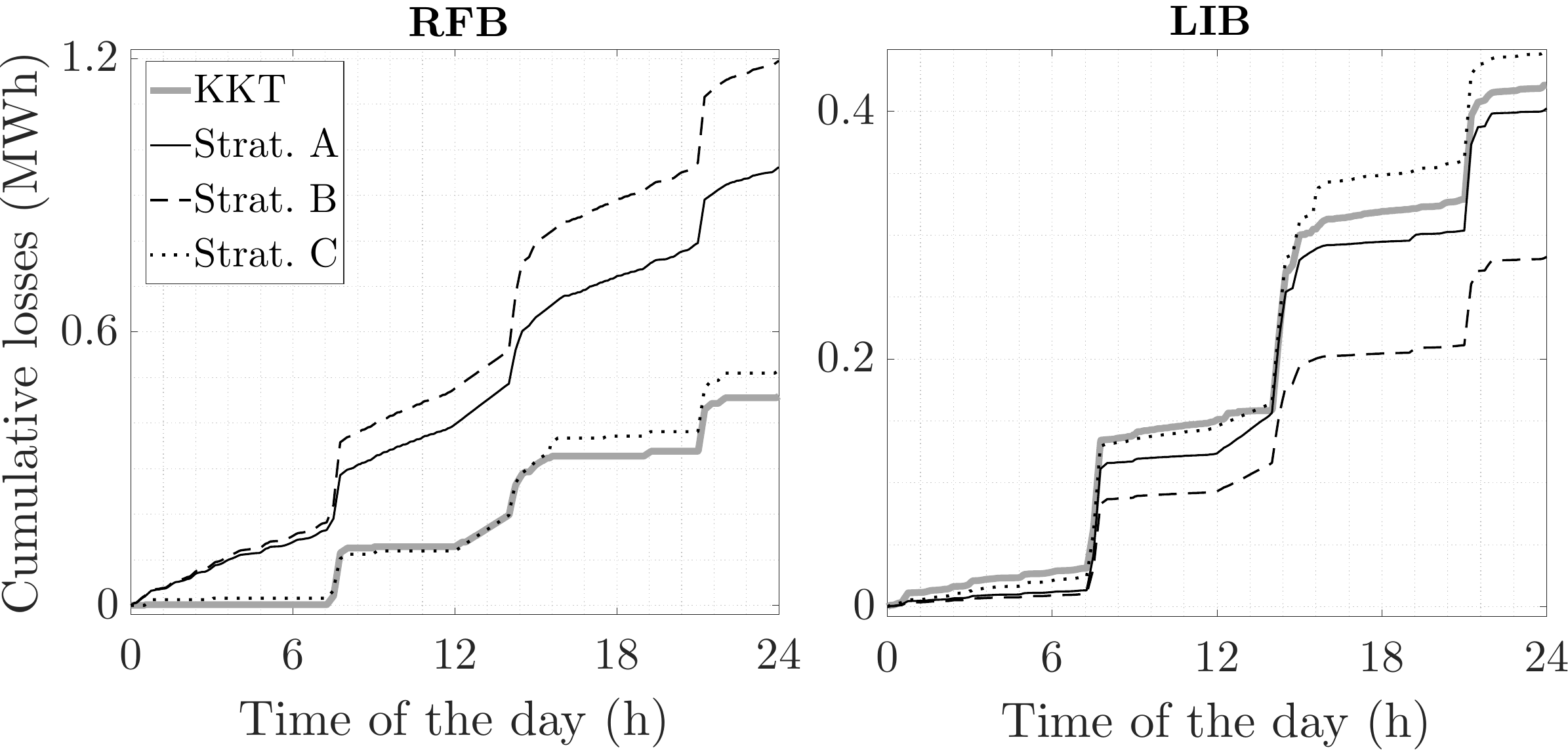}}
\vspace{-0.15cm}
\caption{Cumulative losses of (a) RFB and (b) LIB, for the operation on a full day, for different strategies.}
\label{fig_rev.cumlosses}
\vspace{-0.3cm}
\end{figure}

\begin{figure}[!t]
\centering
{\includegraphics[width=0.93\columnwidth]
{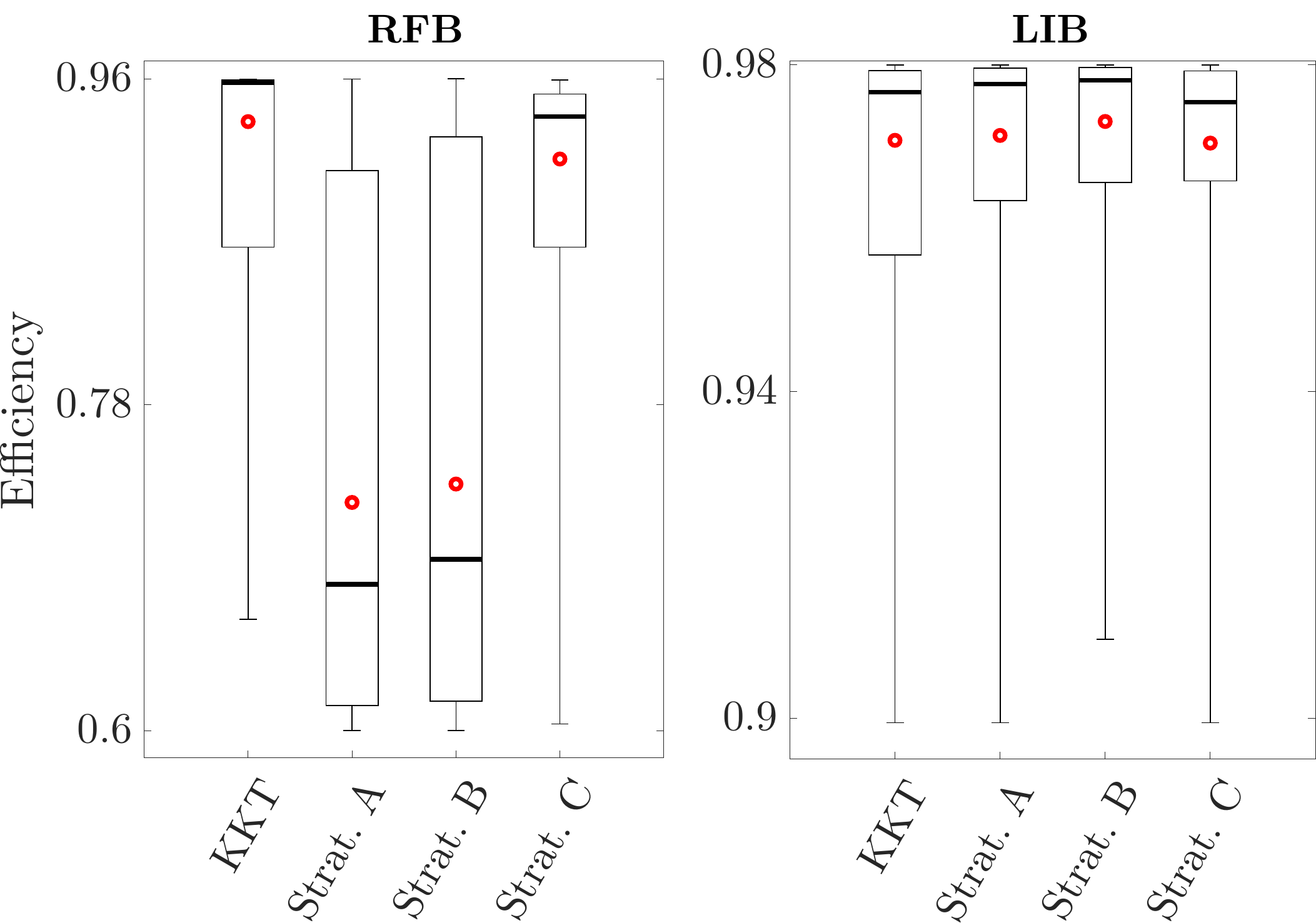}}
\vspace{-0.15cm}
\caption{Efficiency distribution (10th-90th percentile) of the (a) RFB and (b) LIB over a daily operation for different strategies. 
The black bar indicates the median, and the red circle the mean.}
\label{fig_rev.boxplot}
\vspace{-0.3cm}
\end{figure}

To highlight the advantages of the proposed controller, the energy losses are compared for three strategies: 
\begin{itemize}
\item \textit{Strat. A}: Equal sharing of the reference between the units.
\item \textit{Strat. B}: Allocation based on available storage reserves. 

In this case, weighting factors for charging and discharging are calculated based on the SOC of the units~\cite{RiosPEDG}:
\begin{align}
    w_i = \frac{R_i}{\sum_i R_i}, \quad
    R_i = \begin{cases}
    SOC_i - SOC_{\min,i} & \text{(dis.)}, \\
    SOC_{\max,i} - SOC_i & \text{(ch.)}.
    \end{cases}
    \end{align}
    \item \textit{Strat. C}: Power sharing based on rolling horizon optimisation, as done in~\cite{RiosTSTE}.
\end{itemize}
Table~\ref{tab_rev.losses} shows the median efficiency values and the losses over a daily operation, for each storage technology.
Fig.~\ref{fig_rev.cumlosses} shows the cumulative losses and Fig.~\ref{fig_rev.boxplot} shows the efficiency distribution for each strategy.
Clearly, the proposed strategy achieved the highest operational efficiencies, both in terms of mean and median values. 
The better performance than \textit{Strat. C}, which implements an actual optimisation of the power sharing, can be explained by its formulation in \textit{Strat. C} as it uses an approximation of the efficiency curves during operation.
\begin{table}[!t]
    \vspace{-0.4cm}
    \renewcommand{\arraystretch}{1.1}
    \caption{Efficiencies and Losses of the Storage Units for Different Strategies}
    \vspace{-0.15cm}
    \centering
    \begin{tabular}{rrrrr}
    \toprule[0.5pt]
    \textbf{Efficiency} & \textbf{KKT-based} & \textbf{Strat. A} & \textbf{Strat. B} & \textbf{Strat. C} 
    \vspace{0.1cm}\\
    \hline
    RFB & 0.958 & 0.681 & 0.695 & 0.939 \\
    LIB & 0.977 & 0.978 & 0.978 & 0.975 \\ \hline
    \textbf{Losses} (kWh) & & & & 
    \\
    \hline
    RFB & 459.2 & 961.7 & 1195.1 & 516.0 \\
    LIB & 421.3 & 402.4 & 282.5 & 448.2 \\ \hline
    \textbf{Total} & 880.5 & 1364.1 & 1477.6 & 964.2 \\
    \bottomrule[1pt]
    \end{tabular}
    \vspace{-0.1cm}
     \label{tab_rev.losses}
 \end{table}
\subsubsection{PV Mismatch and Reserve Enforcement}
\begin{figure}[!t]
\centering
\centering
\includegraphics[width=\columnwidth]{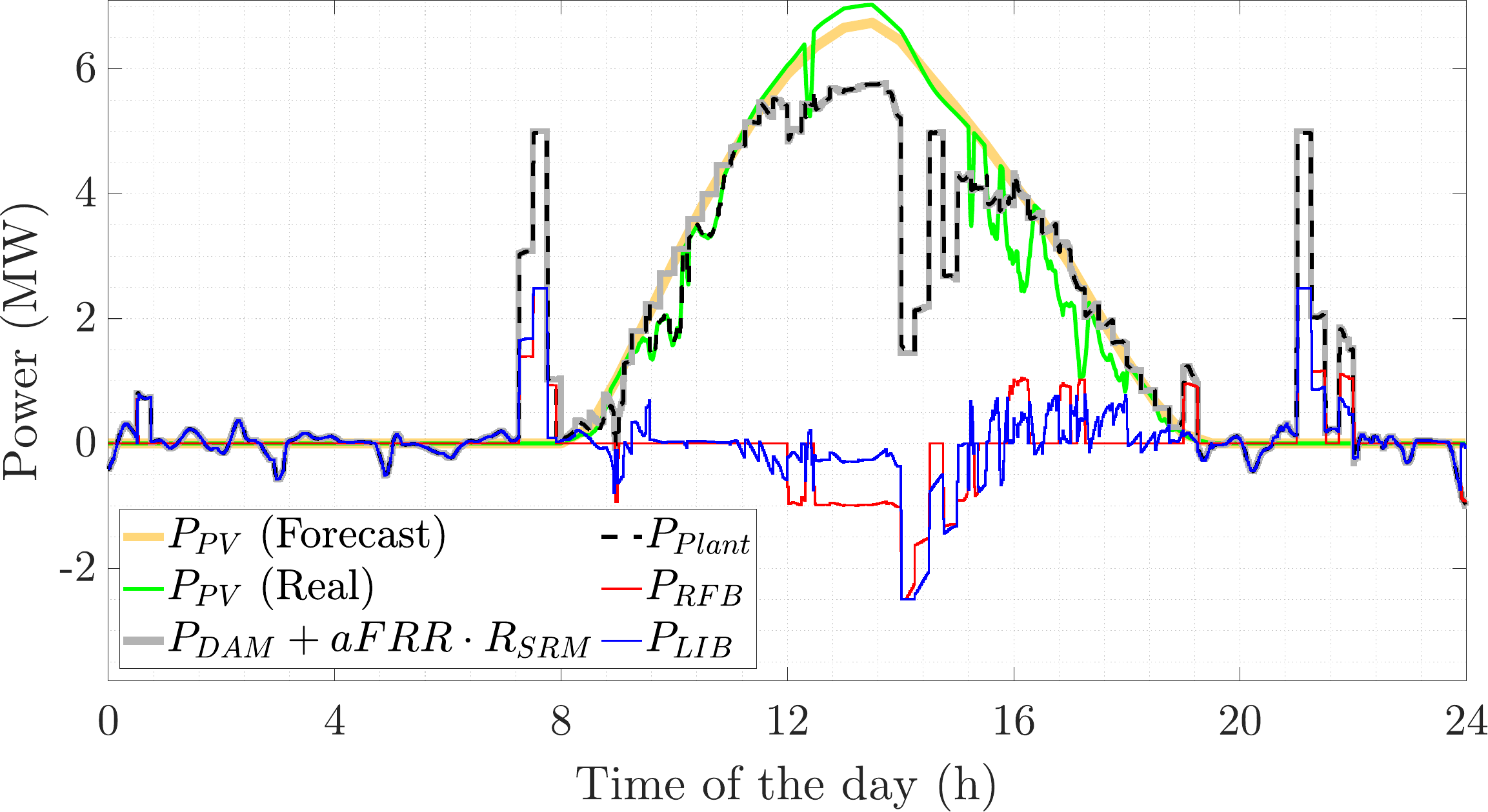}
\vspace{-.4cm}
\caption{Power set-points of the individual storage units and total plant power injection.} 
\vspace{-.2cm}
\label{fig_rev.ProfilesResults_PVreal}
\end{figure}

The operation was performed based on the same schedule obtained in the previous section, but considering a PV generation profile that deviates from the forecasted one.
Since the PV prediction error was expected to have an impact on the available reserves, the ``Reserve enforcement'' block was activated in this section.
Fig.~\ref{fig_rev.ProfilesResults_PVreal} shows the PV forecast and real production, the power profiles of the storage units, the power output of the plant and the committed one.

The KKT-based algorithm was compared only with \textit{Strat. C}.
Similar to the proposed controller, \textit{Strat. C} allowed the HESS to provide energy contracted in the DAM only if the ROC is within the energy limits defined by the contracted reserves.
Through the ``Reserve enforcement'' block, the KKT-based strategy stopped providing energy to the DAM when the reserve limits were violated, but continued providing balancing energy.
This can be seen, e.g., in the period between 8h and 12h, where the power injected by the plant was below the contracted value, mainly due to a shortage of PV generation during those hours compared to the forecast. 

\begin{figure}[!t]
\centering
\centering
\includegraphics[width=\columnwidth]{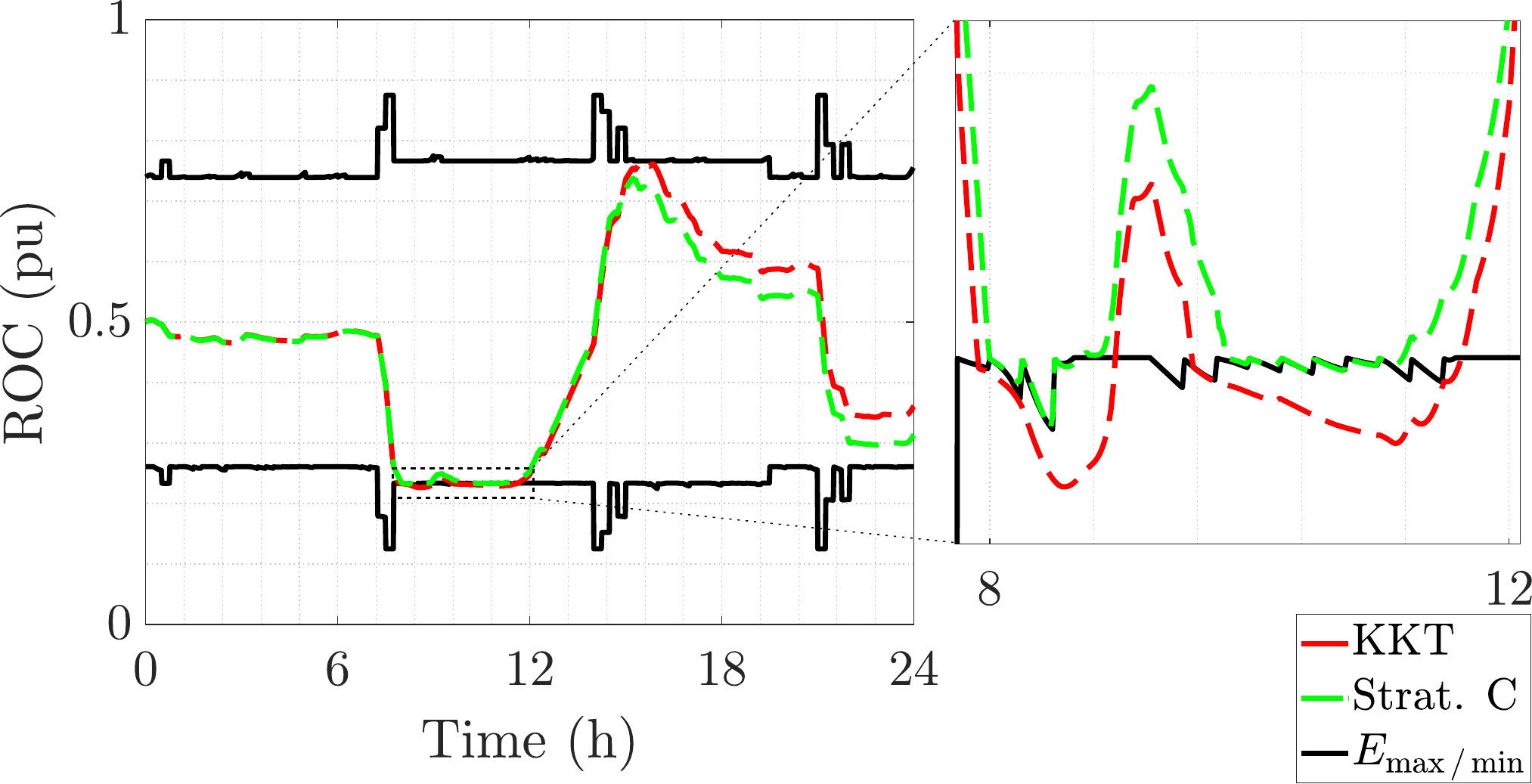}
\vspace{-.16cm}
\caption{Power set-points of the individual storage units and total plant power injection.} 
\vspace{-.2cm}
\label{fig_rev.roc}
\end{figure}

For a more precise analysis, Fig.~\ref{fig_rev.roc} shows the ROC for both strategies and the energy limits for the provision of aFRR service. 
These limits were computed during DAM optimisation such that the plant can maintain the committed reserve capacity for 15 minutes~\cite{entsoe_afrr_2024}.
Since the signals were of one-minute granularity, the minimum and maximum energies ($E_{\min,\max}^{'}$) for each storage unit in the $k$-th step are:
\begin{equation}
\begin{aligned}
{E}_{\min,k}^{'}
&=
SOC_{\min}
+
\frac{15 \cdot {R}_{SRM,k}}{\eta^{+}}, 
\\
{E}_{\max,k}^{'}
&=
SOC_{\max} -\frac{15 \cdot {R}_{SRM,k}}{\eta^{-}}.
\end{aligned}
\label{eq.SOCmaxmin_SRM}
\end{equation}
These limits should be obtained for each storage technology and then aggregated to obtain the ROC limits for the HESS.
During real-time operation, these limits were actually modified by the energy provided for aFRR.
Indeed, when reserve capacity was used to provide balancing energy, the available reserve was reduced without constituting a violation, as the energy was consumed for its intended purpose.
The limits became:
\begin{equation}
\begin{aligned}
{E}_{\min,k}
&=
{E}_{\min,k}^{'} - \sum_{i=k_0}^k aFRR^{+}_i \cdot {R}_{SRM,i} \cdot \Delta t , \\
{E}_{\max,k}
&=
{E}_{\max,k}^{'} + \sum_{i=k_0}^k aFRR^{-}_i \cdot {R}_{SRM,i} \cdot \Delta t,
\end{aligned}
\label{eq.SOCmaxmin_SRM}
\end{equation}
where $\Delta t$ was the one-minute time step of the signals.
The added terms represent the cumulative energy deployed through upward (+) and downward (-) aFRR activations within the current market interval.
Such an interval resets at the start of each new market period, i.e., every 15 minutes; and $k_0$ denotes the first time step of the current interval. 

As shown in Fig.~\ref{fig_rev.roc}, the proposed controller generally respected the energy limits with small deviations.
Deviations occurred because the controller operated without a predictive horizon, preventing it from anticipating and recovering the energy for subsequent market periods when the reserve went below the limit due to aFRR activations.
In contrast, the optimisation-based approach could recover the energy levels needed by the start of each market period, respecting the limits; however, this was achieved at a high computational expense.

Table~\ref{tab_rev.pvreal} shows the median efficiency values and losses during a daily operation for each storage technology, as well as the deviation of the DAM energy.
Both strategies performed similarly, with a slight improvement from the KKT-based one.
\begin{table}[!t]
    \vspace{-0.2cm}
    \renewcommand{\arraystretch}{1.1}
    \caption{Efficiencies and Losses of the Storage Units for Different Strategies}
    \vspace{-0.15cm}
    \centering
    \begin{tabular}{rrr}
    \toprule[0.5pt]
    \textbf{Efficiency} & \textbf{KKT-based} & \textbf{Strat. C} 
    \vspace{0.1cm}\\
    \hline
    RFB & 0.959 & 0.943 \\
    LIB & 0.976 & 0.975 \\ \hline
    \textbf{Losses} (kWh) & & 
    \\
    \hline
    RFB & 483.0 & 593.4  \\
    LIB & 447.0 & 452.6 \\ \hline
    \textbf{Total} & 930.0 & 1046.0 \\
    \bottomrule[1pt]
    \end{tabular}
    \vspace{-0.3cm}
    \label{tab_rev.pvreal}
 \end{table}

\subsection{Real-Time Simulation}
\begin{figure}[!t]
\centering
\includegraphics[width=0.73\columnwidth]{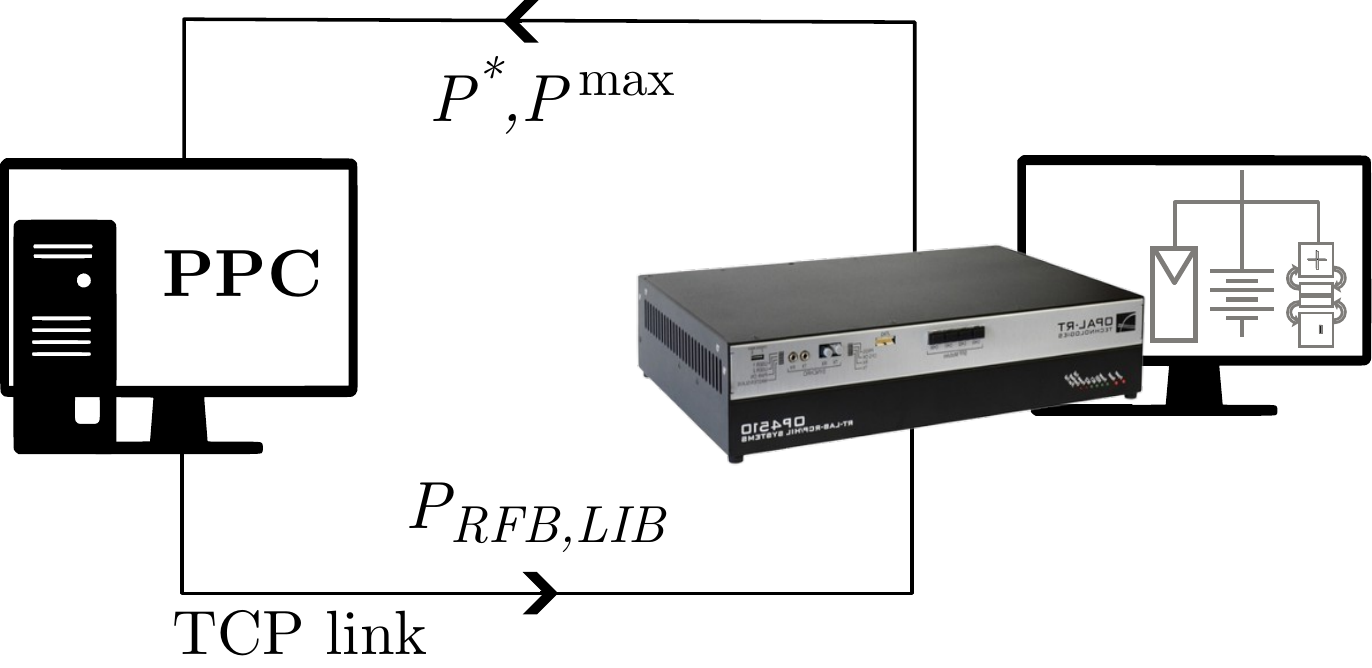}
\vspace{-.2cm}
\caption{Real-time HIL validation setup based on OPAL-RT 4510.} 
\vspace{-.2cm}
\label{fig_rev.rt}
\end{figure}

A real-time simulation was performed to assess the practical feasibility of the proposed algorithm.
Fig.~\ref{fig_rev.rt} shows the real-time Hardware-in-the-Loop (HIL) implementation scheme.
The power plant, including the power devices and their local controllers, was implemented on an OPAL-RT~4510, while the KKT-based algorithm ran on an external computer.
Communication between the two was established via TCP protocol, and the algorithm was executed once every second.

\begin{figure}[!t]
\centering
\centering
\includegraphics[width=0.97\columnwidth]{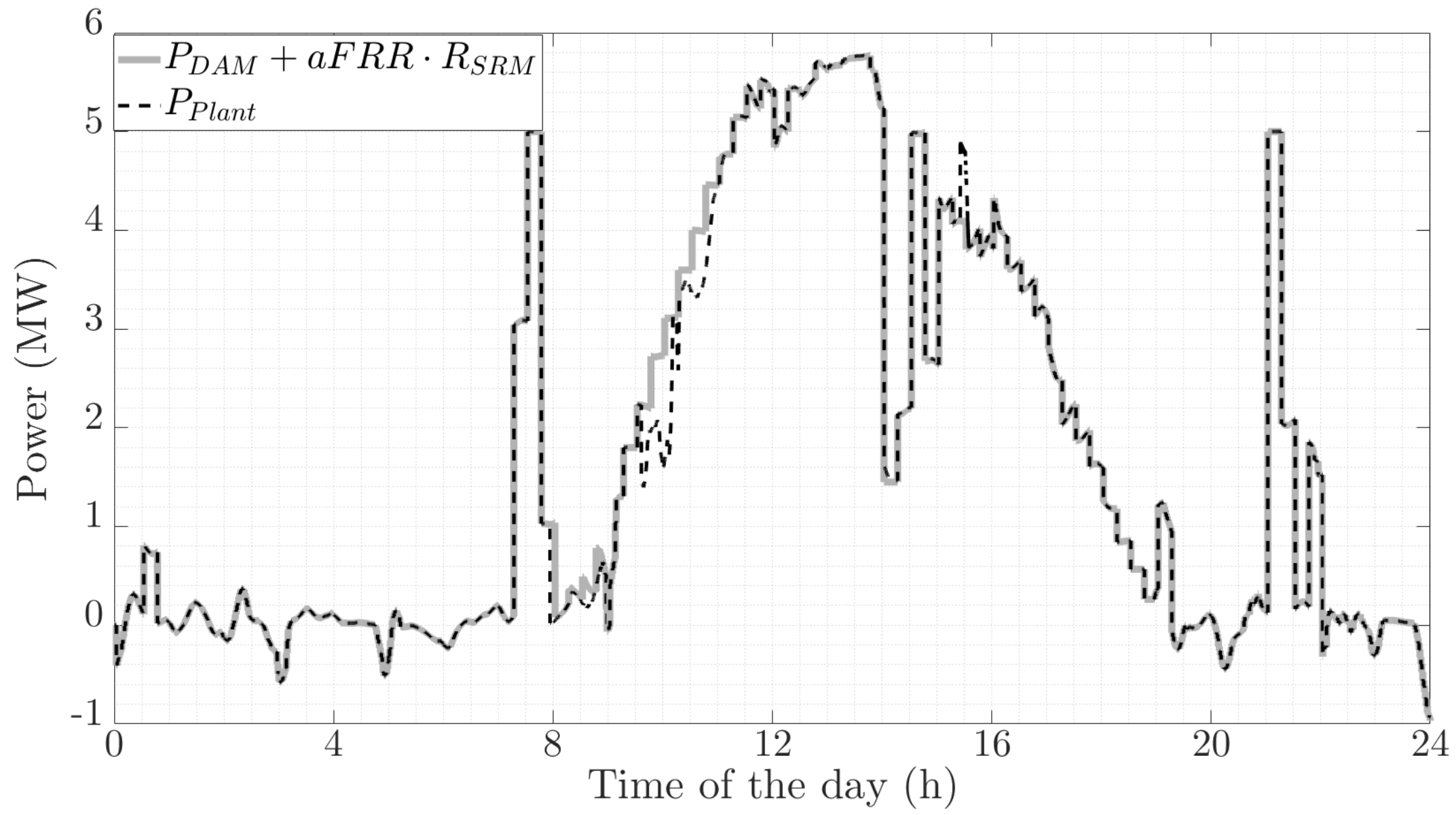}
\vspace{-.25cm}
\caption{Power profiles from real-time simulation.} 
\vspace{-.3cm}
\label{fig_rev.rt_results}
\end{figure}

Fig.~\ref{fig_rev.rt_results} shows the set-point and output of the power plant.
The results show negligible variance compared to those shown in Fig.~\ref{fig_rev.ProfilesResults_PVreal}. 
The mean resolution time of the power sharing algorithm was 6.7~ms, with a standard deviation of 12.0~ms, demonstrating the feasibility of the algorithm and the possibility of executing it more frequently. 
Regarding the performance of the real-time simulator, the mean and maximum step duration for the power system were 8.79~$\mu$s and 9.12~$\mu$s, respectively, while for the control system they were 3.36~$\mu$s and 4.48~$\mu$s
\vspace{-0.1cm}
\subsection{Remarks and Practical Considerations}
Although the proposed controller offers the advantages presented above, some practical aspects and limitations should also be considered:
\begin{enumerate}
\item The proposed approach relies on the convexity of the efficiency curves to ensure the validity of the KKT formulation. 
Depending on the characteristics of the efficiency curve of a technology, or a combination of technologies, this condition may not be satisfied in a straightforward way, potentially requiring a reformulation or an approximation of the curves prior to application.
\item The necessity of maintaining the formulation simple inherently limits the number and complexity of constraints that can be considered.
Nevertheless, some operational constraints can be included in complementary control layers, as in the ``Reserve enforcement'' block. 
\item The myopic nature of the controller prevents it from anticipating the future value of variables and forecast conditions, such as the need to recover reserved energy for subsequent market periods.
This issue could be addressed by imposing more conservative energy limits or by trading in intra-day markets, e.g., to reschedule the reserve capacity or to restore the SOC when it deviates significantly from the expected value.
\end{enumerate}
%
%
\section{Conclusion}
\label{sec.Conclusion}
In this paper, a controller has been proposed that minimises the losses of a hybrid energy storage system (HESS) by considering detailed power-dependent efficiency curves for each storage technology used. 
The power sharing method is based on the analytical verification of the Karush--Kuhn--Tucker (KKT) conditions, avoiding the need for numerical solvers and making the algorithm computationally efficient.
The algorithm was applied to a PV power plant equipped with a HESS composed of a lithium-ion battery (LIB) and a redox-flow battery (RFB).
Since the marginal loss of the RFB is non-monotonic, a convexity power limit $P^c_{RFB}$ is introduced, below which the LIB operating alone is always preferred.
This ensures an unambiguous optimal dispatch.
The plant participates in both the day-ahead market (DAM) and the secondary reserve market (SRM), providing automatic frequency restoration reserve. 
A reserve enforcement block prevents the energy reserved for the SRM from being used for trading energy in the DAM.

The algorithm was initially validated by considering a synthetic profile of the power reference.
It was demonstrated that the controller can track power references while respecting the power and state-of-charge (SOC) limits of each storage device.
Subsequently, the approach was tested in operation during a full day.
Its performance was compared against three benchmark strategies: (i) \textit{Strat. A}: equal sharing between the storage units; (ii) \textit{Strat. B}: allocation based on the resource availability; (iii) \textit{Strat. C}: rolling horizon optimisation. 
The proposed controller achieved the highest operational efficiency and the lowest losses among all methods.
The improvement was particularly evident in the RFB operation, which achieved a median efficiency of 95.8\%. 
The proposed algorithm outperformed even the strategy based on numerical optimisation, though by a small margin, mainly because the latter employs an approximation of the efficiency curves.
The proposed approach leads to the best performance, also in the presence of deviations in PV forecasts. 
Nevertheless, the myopic nature of the controller may lead to small deviations with respect to the energy limits defined by the contracted reserves, whereas strategies based on rolling horizon optimisation strictly respect those limits.
Finally, the controller was tested using a real-time hardware-in-the-loop setup. 
The mean resolution time of the algorithm was 6.7~ms, validating the viability of the algorithm for real-time applications. 

It was demonstrated in this paper that detailed energy storage models should be included in the hybrid power plant dispatch process in order to guarantee improved overall performance of the plant.
At the same time, low computational complexity must be preserved to facilitate its implementation in the provision of ancillary services.
Future work will include an extended formulation to a multi-market framework for the simultaneous provision of multiple services.
\bibliography{IEEEabrv,biblio}

\end{document}